\documentclass[%
 aip,
 amsmath,amssymb,
reprint, 
]{revtex4-1}

\usepackage{graphicx}
\usepackage{dcolumn}
\usepackage{bm}
\usepackage[utf8]{inputenc}
\usepackage[T1]{fontenc}
\usepackage{mathptmx}
\usepackage{etoolbox}

\usepackage{siunitx}
\usepackage{amsmath,amssymb}
\usepackage{xcolor}
\usepackage{hyperref}
\usepackage{booktabs}
\usepackage[version=4]{mhchem}
\usepackage{makecell}

\makeatletter
\def\@email#1#2{%
 \endgroup
 \patchcmd{\titleblock@produce}
  {\frontmatter@RRAPformat}
  {\frontmatter@RRAPformat{\produce@RRAP{*#1\href{mailto:#2}{#2}}}\frontmatter@RRAPformat}
  {}{}
}%
\makeatother
\begin{document}

\author{K. Kakuyanagi}
\affiliation{%
  Basic Research Laboratories, NTT Inc., 3-1 Morinosato-Wakamiya, Atsugi, Kanagawa, 243-0198, Japan
}%
\author{N. Teran}
\affiliation{%
  Basic Research Laboratories, NTT Inc., 3-1 Morinosato-Wakamiya, Atsugi, Kanagawa, 243-0198, Japan
}%
\affiliation{%
  School of Chemistry, University of Edinburgh, David Brewster Road, Edinburgh, EH9 3FJ, Scotland, UK
}%
\author{H. Toida}
\affiliation{%
  Basic Research Laboratories, NTT Inc., 3-1 Morinosato-Wakamiya, Atsugi, Kanagawa, 243-0198, Japan
}%
\author{S. Saito}
\affiliation{%
  Basic Research Laboratories, NTT Inc., 3-1 Morinosato-Wakamiya, Atsugi, Kanagawa, 243-0198, Japan
}%
\email{kosuke.kakuyanagi@ntt.com}

\date{\today}

\begin{abstract}
  Achieving uniform critical current across Josephson junctions is essential for the large-scale integration of superconducting quantum circuits.
  In this work, we statistically analyzed the variation of the critical current of \ce{Al/AlO_x/Al} junctions using room-temperature tunnel resistance statistics,
  and identified the dominant contribution among the modeled sources of the variation based on their dependence on geometry and deposition conditions of junctions.
  Our model-based analysis reveals that fluctuations in the Al film thickness play the dominant role among the modeled contributing factors.
  Based on this analysis, we found that, in Dolan-bridge double-angle deposition, adopting a deposition angle of \SI{30}{\degree} for bilayer junctions significantly improves uniformity, yielding a relative standard deviation of \SI{1.2}{\percent} (\SI{0.5}{\percent}) across a \SI{9.75}{mm} (\SI{1.5}{mm}) square region.
\end{abstract}

\title{Geometric dependence of critical-current variation in \ce{Al/AlO_x/Al} Josephson junctions: a model-based analysis}

\maketitle

Al/AlO\textsubscript{x}/Al Josephson junctions fabricated by the Dolan-bridge technique \cite{10.1063/1.89690} and Manhattan patterning \cite{Potts2001, 10.1116/1.4722982} are widely used in various quantum devices, including superconducting qubits \cite{Nakamura1999}. 
The critical current $I_\mathrm{C}$ determines the Josephson energy $E_\mathrm{J}={\Phi_0 I_\mathrm{C}/ 2\pi}$ and the Josephson inductance $L_\mathrm{J}={\Phi_0/ 2\pi I_\mathrm{C}}$; therefore, its variation directly limits circuit performance.
Depending on the application, different levels of uniformity are required. 
For example, in transmon-based \cite{PhysRevA.76.042319} quantum processors \cite{Arute2019}, uniformity on the order of \SIrange{0.5}{2}{\percent} is required to avoid frequency collisions \cite{8614500}. 
Similarly, high uniformity is also essential for multi-junction devices such as Josephson traveling-wave parametric amplifiers (JTWPA) \cite{10.1063/1.4922348,doi:10.1126/science.aaa8525,Peatain_2023}.

To improve the uniformity of Josephson junctions, various process optimizations have been reported \cite{Kreikebaum_2020, Takahashi_2023, Moskalev2023, Pishchimova2023, Zheng2023}.
These studies demonstrated that improved control of lithographic patterning, shadow-evaporation geometry, and oxidation conditions significantly reduces junction-area fluctuations and tunnel-barrier variations, leading to a few-percent-level dispersion in the 
critical current at the wafer scale. 
However, the microscopic mechanisms that determine the effective junction area and tunnel-barrier thickness are not yet fully understood. 
As a result, the fundamental origins of device-to-device variations remain unresolved.

In this work, motivated by these issues, we constructed a model of Dolan-bridge Josephson junctions based on the viewpoint that variations in the critical current arise from fluctuations in Al-film thickness, junction area, and edge roughness, each exhibiting distinct dependencies on geometry.
To investigate the origins of these variations, we fabricated test junctions with systematically varied geometries and measured their room-temperature resistance.
Our analysis shows that film-thickness-related fluctuations are the dominant contribution among the modeled sources of critical-current variations and provides quantitative design guidelines for improving junction uniformity in large-scale circuits.

The critical current of a superconducting tunnel junction $I_\mathrm{C}$ can be evaluated from the normal-state tunnel resistance $R_\mathrm{N}$ using the Ambegaokar--Baratoff relation \cite{PhysRevLett.10.486}:
\begin{equation}
  I_\mathrm{C} R_\mathrm{N} = \frac{\pi \Delta}{2e} \tanh\!\left(\frac{\Delta}{2k_\mathrm{B}T}\right),
\end{equation}
where $\Delta$ is the superconducting gap,
$e$ is the elementary charge,
$k_\mathrm{B}$ is the Boltzmann constant,
and $T$ is the temperature.
Assuming that $\Delta$ is constant and the hyperbolic-tangent term approaches unity at the operating temperature of quantum devices,
the product of $I_\mathrm{C}$ and $R_\mathrm{N}$ can be treated as a constant.
Thus, the variation in the critical current can be related to that of the normal-state conductance $G_\mathrm{N}=1/R_\mathrm{N}$.
From this relationship, the variation in conductance $\delta G_\mathrm{N}$ can be expressed as
\begin{equation}
  \delta G_\mathrm{N} = \frac{\mathrm{d}G_\mathrm{N}}{\mathrm{d}R_\mathrm{N}}\,\delta R_\mathrm{N} = -\frac{1}{R_\mathrm{N}^2}\,\delta R_\mathrm{N}.
\end{equation}
Assuming the variations are much smaller than their average values, and from the definition of variance $\mathrm{Var}(x)=
\left\langle \left(\delta x\right)^2\right\rangle$,
we obtain the relative variances
\begin{equation}
  \frac{\mathrm{Var}(I_\mathrm{C})}{\langle I_\mathrm{C}\rangle^2}
  =\frac{\mathrm{Var}(G_\mathrm{N})}{\langle G_\mathrm{N}\rangle^2}
  \approx \frac{\mathrm{Var}(R_\mathrm{N})}{\langle R_\mathrm{N}\rangle^2}.
\end{equation}
Therefore, the room-temperature distribution of the normal-state resistance provides a direct estimate of the relative variations in both conductance and critical current.

To isolate short-range, intrinsic process-induced variation from long-range spatial fluctuations, we introduce the semivariogram and its normalized form instead of relative standard deviation
\begin{align}
  \gamma(\mathbf{r})         & = \frac{1}{2}\left\langle \left[R_\mathrm{N}(\mathbf{r}_0+ \mathbf{r}) - R_\mathrm{N}(\mathbf{r}_0)\right]^2 \right\rangle_{\mathbf{r}_0}, \\
  \tilde{\gamma}(\mathbf{r}) & = \frac{\gamma(\mathbf{r})}{\langle R_\mathrm{N} \rangle^2},
\end{align}
where $\mathbf{r}$ is the displacement vector between two positions.
It is worth noting that, under the assumption of second-order spatial stationarity, the semivariogram can be expressed as the difference between the variance and the spatial covariance,
\begin{equation}
  \gamma(\mathbf{r})
  = \mathrm{Var}(R_\mathrm{N})
  - \mathrm{Cov}\!\left(
    R_\mathrm{N}(\mathbf{r}_0+\mathbf{r}),
    R_\mathrm{N}(\mathbf{r}_0)
  \right).
\end{equation}
Consequently, when spatial correlations vanish, $\gamma(\mathbf{r})$ reduces to $\mathrm{Var}(R_\mathrm{N})$.
By setting $\mathbf{r}$ to the nearest-neighbor spacing of the junction array, we obtain a metric that predominantly quantifies intrinsic, process-induced variation between neighboring junctions.
In the following, we refer to $\sqrt{\tilde{\gamma}}$ as the normalized RMS (root-mean-square) variation.

The variation of conductance is modeled as the sum of random fluctuations that are decomposed into contributions with different dependence on geometry and process parameters.
In this model, we focus on intrinsic junction-to-junction variations.
Long-range spatial variations, such as resist thickness undulations introduced during spin coating and smooth wafer-scale Al-thickness gradients originating from the source-to-sample distance or deposition angle, are largely suppressed by evaluating the semivariogram at the nearest-neighbor spacing.
The film-thickness variation considered below therefore refers to residual local variations in the effective Al deposition geometry in the junction-forming region.
\begin{equation}
  \delta G_{\mathrm{N}} =
  \delta G_0
  + \delta G_S(S)
  + \delta G_L(L)
  + \delta G_d(\theta,N_\mathrm{L},W)
  \label{eqS:additive}
\end{equation}
Here, $\delta G_0$ denotes geometry-independent fluctuations,
$\delta G_S$ and $\delta G_L$ represent structural fluctuations that depend on area $S$ and edge length $L$, respectively,
and $\delta G_d$ represents fluctuations induced by film-thickness variations,
with a magnitude that depends on the deposition angle $\theta$, the number of layers $N_\mathrm{L}$, and the junction width $W$.
We assume that these fluctuation components are statistically independent, as they originate from distinct physical mechanisms and spatial scales within the fabrication process, such as lithography and deposition.
The variances are modeled as follows:
\begin{align}
  \label{eqS:model_add}
  \mathrm{Var}(G_\mathrm{N}) &= \mathrm{Var}(G_0) + \mathrm{Var}(G_S) + \mathrm{Var}(G_L) + \mathrm{Var}(G_d), \\
  \mathrm{Var}(G_0) & = c_0,                             \\
  \mathrm{Var}(G_S) & = c_S S + c_{S^2} S^2,             \\
  \mathrm{Var}(G_L) & = c_L L + c_{L^2} L^2,             \\
  \mathrm{Var}(G_d) & = c_df^2(\theta, N_\mathrm{L})W^2.
\end{align}
Here, the linear term in $S$ arises from fluctuations of the oxide tunnel-barrier thickness or barrier properties that occur at a microscopic scale within the junction.
The quadratic term reflects long-range spatial fluctuations that vary between junctions, such as fluctuations in the oxide tunnel-barrier thickness.
Similarly for $L$, the linear term corresponds to short-range line-edge roughness along the junction edge, whereas the quadratic term corresponds to long-range spatial fluctuations of pattern shape and resist thickness.
For the thickness-related contribution, the fluctuation in junction area induced by film-thickness variations is proportional to the junction width $W$, and therefore the corresponding variance scales with $W^2$.
The contributions associated with $S$ and $L$ are distinguished through their different dependences on junction geometry when the junction size is varied.
The thickness-related term is further separated by its dependence on the number of layers $N_\mathrm{L}$ through the geometric factor $f(\theta,N_\mathrm{L})$.

\begin{figure}[t]
  \centering
  \includegraphics[width=0.9\linewidth]{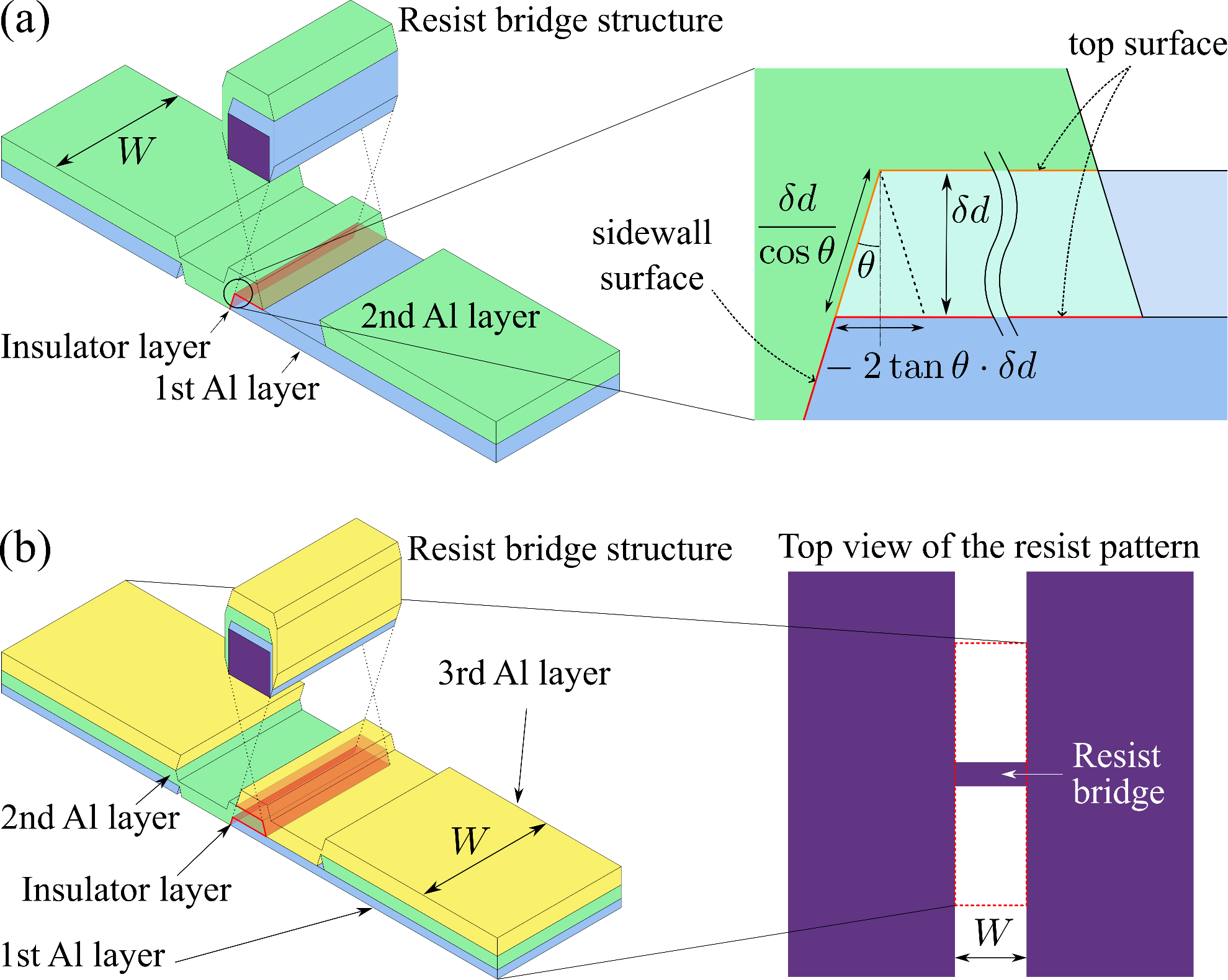}
  \caption{\label{fig:S1}
(a) Cross-sectional schematic of the \ce{Al/AlO_x/Al} bilayer junction formed by the Dolan-bridge (purple) double-angle deposition.
The blue (green) layer represents the first (second) deposited Al film.
The surfaces highlighted in red correspond to the junction interfaces with area $S$, and the perimeter of these surfaces corresponds to the junction-edge length $L$.
The magnified schematic illustrates the relationship among the deposition angle $\theta$, the film-thickness fluctuation $\delta d$, and the junction interfaces (red and orange).
The sidewall surface and the top surface constitute the junction interfaces.
When the film thickness increases by $\delta d$, the junction interface shifts from the line shown in red to the line shown in orange.
(b) Cross-sectional schematic of the trilayer junction.
The yellow layer represents the Al film deposited in the third evaporation.
The right panel shows the top view of the resist pattern.
}
\end{figure}

The geometric origin of the dependence on deposition angle is illustrated in Fig.~\ref{fig:S1}.
During angled deposition, the junction is formed both on the top and sidewall surfaces.
The junction interface consists of both the top surface and the sidewall surface (Fig.~\ref{fig:S1}(a)). When the film thickness changes, these two contributions change with different geometric factors, resulting in different sensitivities to film-thickness fluctuations.
As a result, the contribution of film-thickness fluctuations to the critical-current variation depends on the deposition angle $\theta$.
This effect is formulated as the dimensionless geometric factor:
\begin{equation}
  f^2(\theta, N_\mathrm{L})=
  \frac{N_\mathrm{L}-1}{\cos^2\theta}
  \left[
    2N_\mathrm{L}\sin\theta
    \left(\frac{2N_\mathrm{L}-1}{3}\sin\theta-1\right)
    +1
    \right],
  \label{eqS:fNL}
\end{equation}
whose derivation is provided in the Supplemental Material.
In particular, for bilayer junctions ($N_\mathrm{L}=2$), the geometric factor is calculated
\begin{equation}
f^2(\theta,2) = \left(\frac{1}{\cos\theta} - 2 \tan\theta \right)^2.
\end{equation}
The origin of each term can be understood as follows.
As illustrated in the magnified schematic in Fig.~\ref{fig:S1}(a), an increase in film thickness increases the contribution from the sidewall surface in proportion to $1/\cos\theta$, while it decreases the contribution from the top surface in proportion to $2\tan\theta$.
This factor vanishes at $\theta = \SI{30}{\degree}$, whereas no such null angle exists for $N_\mathrm{L}\ge 3$.

To verify the dependence of the model on geometry and process conditions,
approximately 4000 \ce{Al/AlO_x/Al} Josephson junctions were fabricated in a \SI{13}{\milli\meter} square region of a \SI{17}{\milli\meter} square Si chip by the Dolan-bridge technique \cite{10.1063/1.89690}.
Si chips were coated with a \SI{300}{\nano\meter} PMGI resist layer followed by a \SI{100}{\nano\meter} ZEP resist layer.
The designed Dolan-bridge width was \SI{100}{\nano\meter} for all devices.
Patterns, corrected for the proximity effect using BEAMER software (GenISys GmbH), were written by electron-beam lithography (JEOL JBX-8100FS) and developed with butyl acetate and PMGI 101A developers.
The Al films were deposited using a PLASSYS MEB550S2-I UHV evaporator with deposition angles of $\pm \SI{17}{\degree}$ or $\pm \SI{30}{\degree}$. 
The thickness of the deposited Al layers ranged from approximately 10 to 70 nm. 
For the experiments investigating the dependence on junction geometry, the bilayer junctions had thicknesses of 40 and 50 nm, while the trilayer junctions had thicknesses of 10, 20, and 60 nm. For the experiments investigating the dependence on deposition angle, bilayer junctions were used, with layer thicknesses of 20 and 70 nm.
Dynamic oxidation was performed between the depositions at an oxygen pressure of \SI{10}{\milli Torr} for \SI{30}{\minute}.
During oxidation, the sample holder was rotated at five revolutions per minute to enhance spatial uniformity, following a fabrication procedure similar to that described in Ref.~\onlinecite{Kreikebaum_2020}.

The four-terminal resistance of the junctions was measured automatically at room temperature using a probe station.
Outliers deviating by more than $\pm \SI{20}{\percent}$ from the median value were treated as open or shorted junctions and excluded from the analysis.
This threshold is much larger than the intrinsic variation discussed below, and is used only to remove clearly defective junctions.

To examine the geometric dependence, we fabricated and measured test junctions with four different junction widths at deposition angles of $\pm \SI{17}{\degree}$.
To ensure identical process conditions, junctions with each width were fabricated in separate \SI{6.2}{\milli\meter}-square regions on the same chip, each containing 1024 test junctions.
Additionally, bilayer junctions and trilayer junctions, as shown in Fig.~\ref{fig:S1}(b), were fabricated to evaluate the effect of the number of Al layers $N_\mathrm{L}$.
Note that the multilayer junction considered here does not correspond to Josephson junctions connected in series; rather, the successive deposition layers do not form electrically distinct junctions but instead contribute in parallel to a single effective junction interface with increased area.
\begin{figure} 
  \includegraphics[width=0.75\linewidth]{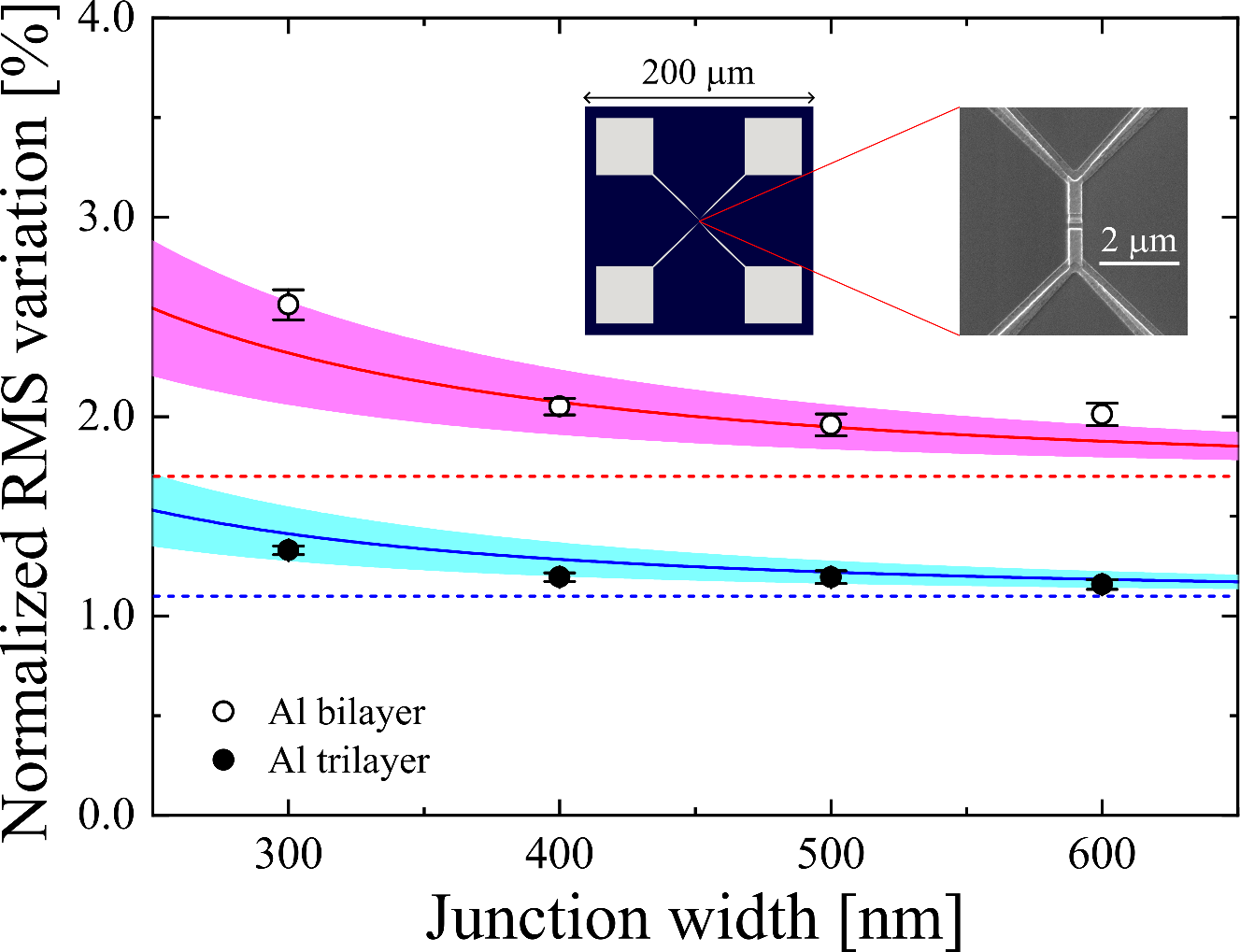}
  \caption{\label{fig:fig001}
    Geometry dependence of the measured normalized RMS variation and the corresponding model fitting results.
    White (black) circles represent the measured data for the bilayer (trilayer) junctions, and the error bars are estimated from the unbiased sample variance.
    The red (blue) solid line represents the model fitting results for the bilayer (trilayer) junctions.
    The dashed horizontal lines indicate the asymptotic values of the normalized RMS variation for the bilayer and trilayer junctions, which correspond to the contribution from film-thickness fluctuations.
    The magenta (cyan) shaded region indicates the uncertainty of the fitted curve for the bilayer (trilayer) junctions, calculated from the standard errors of the fitting parameters.
    Inset (top center): pattern of the unit cell of a test junction.
    For four-terminal measurements, each test junction is connected to four square electrodes with a side length of \SI{50}{\micro\meter}.
    Inset (top right): a scanning electron microscope image of the test junction.
  }
\end{figure}

Figure~\ref{fig:fig001} shows the normalized RMS variation as a function of junction width, together with the fitting results using the model [Eq.~(\ref{eqS:model_add})].
Here, we fit the variance of the normalized resistance variation to the model. 
The variance is obtained as the square of the experimentally measured normalized RMS variation, and the model describes the intrinsic process-induced variance.
The combined dataset consists of eight measured values obtained by varying the junction width $W$ and the number of deposited layers $N_\mathrm{L}$, whereas the model contains six non-negative fitting coefficients.
In addition, the model terms have distinct functional dependences on $W$ and $N_{\rm L}$, and the corresponding basis functions are linearly independent for the measured set of geometries.
Therefore, the fitting problem is sufficiently constrained by the present dataset and yields a unique least-squares solution.
The fitting parameters were determined from the geometry dependence of the data using a non-negativity-constrained least-squares fit, so that all variance components remain physically meaningful.
The fitting results indicate that the geometry-independent term $c_0$ and the film-thickness-related term $c_d$ constitute the dominant contributions to the total variance.
In the constrained fit, the optimal solution satisfying the non-negativity constraints sets the area- and edge-related coefficients to zero, indicating that these terms are not required to explain the present dataset.
The lower normalized RMS variation in trilayer junctions is consistent with the statistical averaging of film-thickness fluctuations across the layers, as the fluctuations in each layer are independent.

Table~\ref{tabS:decomp} summarizes the estimated contributions of geometry-independent and film-thickness-related terms for different junction geometries.
\begin{table}[h]
  \centering
  \caption{
    Summary of the measured normalized RMS variation of the resistance, the corresponding model-fitted relative standard deviation, and the estimated contributions of the geometry-independent and film-thickness-related terms for different junction geometries.
    The thickness-induced terms are calculated for each $N_\mathrm{L}$ from the fitting curve of $W$ dependence, thus the same values appear for the same $N_\mathrm{L}$.
    }
  \sisetup{table-format=1.2}
  \begin{tabular}{@{}cccccc@{}}
    \toprule
    \makecell{$W$ \\(nm)}   & \makecell{$N_\mathrm{L}$\\\ } & \makecell{Measured\\$\sqrt{\tilde{\gamma}}$ }        & \makecell{Model\\$\sqrt{\mathrm{Var}(G_\mathrm{N})}/G_\mathrm{N}$}        & \makecell{Independent\\$\sqrt{\mathrm{Var}(G_0)}/G_\mathrm{N}$}          & \makecell{Thickness \\$\sqrt{\mathrm{Var}(G_d)}/G_\mathrm{N}$}           \\
    \midrule
    $300$ & 2 & $2.56\!\pm\! \SI{0.08}{\percent}$ & $2.32\!\pm\! \SI{0.21}{\percent}$ & $1.57\!\pm\! \SI{0.61}{\percent}$ & $1.70\!\pm\! \SI{0.18}{\percent}$ \\
    $400$ & 2 & $2.05\!\pm\! \SI{0.04}{\percent}$ & $2.07\!\pm\! \SI{0.17}{\percent}$ & $1.18\!\pm\! \SI{0.46}{\percent}$ & $1.70\!\pm\! \SI{0.18}{\percent}$ \\
    $500$ & 2 & $1.96\!\pm\! \SI{0.06}{\percent}$ & $1.95\!\pm\! \SI{0.11}{\percent}$ & $0.94\!\pm\! \SI{0.37}{\percent}$ & $1.70\!\pm\! \SI{0.18}{\percent}$ \\
    $600$ & 2 & $2.01\!\pm\! \SI{0.06}{\percent}$ & $1.88\!\pm\! \SI{0.07}{\percent}$ & $0.79\!\pm\! \SI{0.31}{\percent}$ & $1.70\!\pm\! \SI{0.18}{\percent}$ \\
    $300$ & 3 & $1.33\!\pm\! \SI{0.02}{\percent}$ & $1.41\!\pm\! \SI{0.15}{\percent}$ & $0.89\!\pm\! \SI{0.35}{\percent}$ & $1.10\!\pm\! \SI{0.12}{\percent}$ \\
    $400$ & 3 & $1.20\!\pm\! \SI{0.02}{\percent}$ & $1.28\!\pm\! \SI{0.10}{\percent}$ & $0.67\!\pm\! \SI{0.26}{\percent}$ & $1.10\!\pm\! \SI{0.12}{\percent}$ \\
    $500$ & 3 & $1.19\!\pm\! \SI{0.03}{\percent}$ & $1.22\!\pm\! \SI{0.06}{\percent}$ & $0.53\!\pm\! \SI{0.21}{\percent}$ & $1.10\!\pm\! \SI{0.12}{\percent}$ \\
    $600$ & 3 & $1.16\!\pm\! \SI{0.02}{\percent}$ & $1.18\!\pm\! \SI{0.04}{\percent}$ & $0.44\!\pm\! \SI{0.17}{\percent}$ & $1.10\!\pm\! \SI{0.12}{\percent}$ \\
    \hline
    \bottomrule
  \end{tabular}
  \label{tabS:decomp}
\end{table}
Based on the quantitative decomposition summarized in Table~\ref{tabS:decomp}, these results clearly show that the contribution from film-thickness fluctuations $\sqrt{{\rm Var}(G_d)}/G_{\rm N}$ is the dominant component among the modeled sources of the variation.
For the bilayer junctions deposited at \SI{17}{\degree}, the fitted film-thickness-related contribution is approximately $1.7\%$.
Using the designed Dolan-bridge width of \SI{100}{nm} and the bilayer geometric factor $f(\SI{17}{\degree},2)=1/\cos\SI{17}{\degree}-2\tan\SI{17}{\degree}$, this value corresponds to an effective Al-thickness variation of approximately \SI{4}{nm}.
This value is not unrealistic as a local thickness variation in the effective junction-forming region.
Moreover, since the standard deviation associated with the geometry-independent term $\sqrt{{\rm Var}(G_0)}$ remains constant, its relative standard deviation $\sqrt{{\rm Var}(G_0)}/G_{\rm N}$ decreases as the junction area increases.
This finding indicates that increasing the junction width is an effective way to suppress variations in practical fabrication processes.

We further examine the effect of deposition angle. 
For this purpose, bilayer junctions with a width of \SI{300}{nm} were fabricated at deposition angles of \SI{17}{\degree} and \SI{30}{\degree}. 
The samples fabricated at \SI{30}{\degree} exhibit a substantially lower normalized RMS variation of $1.15\pm0.03\%$ compared with $1.92\pm0.03\%$ for those fabricated at \SI{17}{\degree}, corresponding to an overall reduction by a factor of $1.67\pm0.04$. 
This improvement, however, cannot be explained solely by the increase in junction area.
The increase in junction area resulting from changing the deposition angle from \SI{17}{\degree} to \SI{30}{\degree}, estimated from the average $R_\mathrm{N}$, is $2.02\pm0.09$, which corresponds to a reduction factor of $1.42\pm0.03$ in the normalized RMS variation.
Assuming the same junction area as that obtained with the \SI{30}{\degree} deposition but a deposition angle of \SI{17}{\degree}, the expected normalized RMS variation is $1.35\pm0.04\%$.
The experimentally observed value at \SI{30}{\degree} ($1.15\pm0.03\%$) is significantly lower than this expectation, indicating an additional suppression of film-thickness-induced contributions to the critical-current variation due to the deposition angle, consistent with the geometric factor in Eq.~(\ref{eqS:fNL}).
This deposition-angle dependence also helps distinguish the film-thickness-related contribution from possible short-range resist-thickness variations.
A local resist-thickness fluctuation would not be expected to follow the geometric factor $f(\theta,N_\mathrm{L})$, whereas the observed additional suppression at \SI{30}{\degree} is consistent with the vanishing of the bilayer thickness-related factor.

\begin{figure}
  \includegraphics[width=1.0\linewidth]{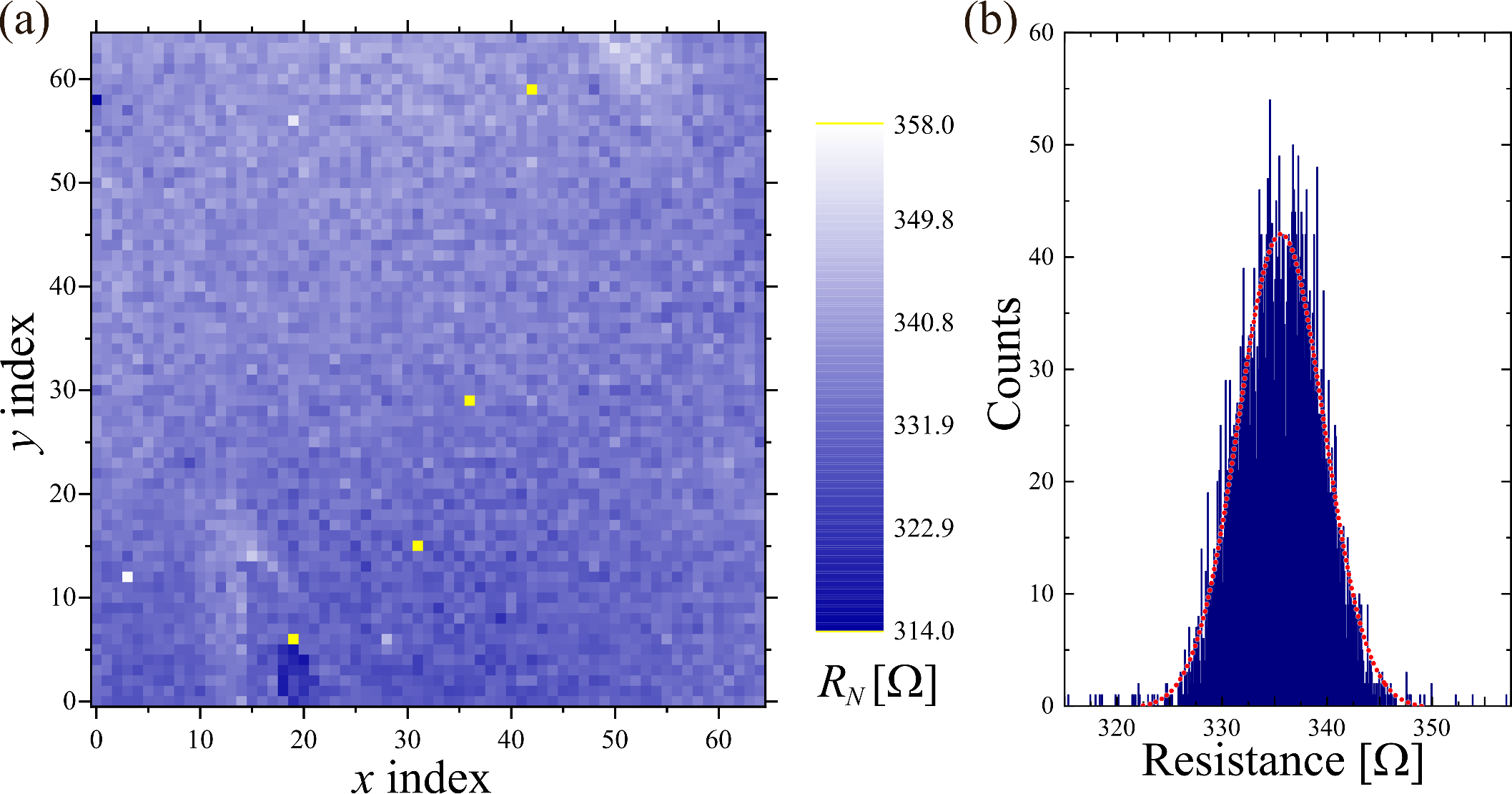}
  \caption{\label{fig:fig002}
    Results of room-temperature resistance measurements for the \SI{9.75}{mm} square junction array.
    (a) Two-dimensional map of the measured resistance. The four yellow points correspond to open or shorted junctions, which are excluded from the statistical analysis.
    (b) Histogram of the resistance values for all valid junctions. The red dashed line is a Gaussian curve shown as a guide to the eye.
  }
\end{figure}
Based on the preceding experiments and analyses, we found that employing a bilayer structure with a \SI{30}{\degree} deposition angle is effective for suppressing the contribution of film-thickness fluctuations, and that increasing the junction area is effective for reducing the variation originating from geometry-independent terms.
Motivated by these findings, we fabricated and evaluated bilayer Josephson junctions using a \SI{30}{\degree} deposition angle, with the junction width set to \SI{600}{nm} to further enhance uniformity.

Maintaining a uniform resist thickness is crucial in the Dolan-bridge process.
To avoid the resist beads near the edge of the \SI{17}{mm} square chip, the unit-cell size of the test junctions was reduced from \SIrange{200}{150}{\micro\meter}, and junctions were fabricated within a \SI{9.75}{mm} square region.

Figure~\ref{fig:fig002} shows the results of the resistance measurements. The yield exceeded \SI {99.9}{\percent} [Fig.~\ref{fig:fig002}(a)].
As shown in Fig.~\ref{fig:fig002}(b), the histogram of the room-temperature resistance exhibits a nearly symmetric distribution, with a mean value of \SI{335.5}{\ohm}, a standard deviation of \SI{3.99}{\ohm}, and a relative standard deviation of \SI{1.19}{\percent} (4221 valid samples). This level of relative variation is comparable to the lowest values reported in the literature evaluating the variation of sparsely arrayed qubits in \SI{10}{\milli \meter} square regions \cite{Kreikebaum_2020,Takahashi_2023}.

\begin{figure*}[!t]
  \centering
  \includegraphics[width=0.9\linewidth]{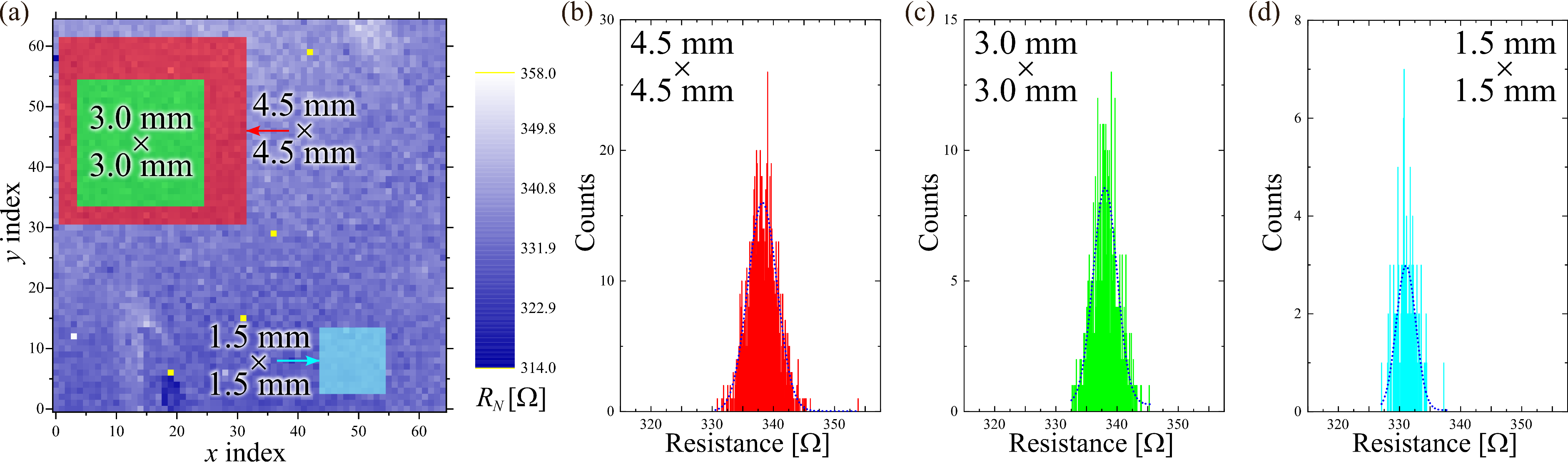}
  \caption{
    (a) Schematic of three representative regions: \SI{4.5}{\milli\meter} square (red), \SI{3.0}{\milli\meter} square (green), and \SI{1.5}{\milli\meter} square (cyan).
    The dataset is the same as that in Fig.~\ref{fig:fig002}. These regions are chosen to minimize standard deviations.
    (b--d) Corresponding histograms of the measured resistance values in these three regions.
    The dashed lines are Gaussian curves shown as guides to the eye.
  }
  \label{figS:sdshape}
\end{figure*}
In addition, spatially correlated nonuniformities in the resistance are observed in the lower-left and upper-right regions of the chip [Fig.~\ref{fig:fig002}(a)], which are likely caused by resist-thickness undulations introduced during the spin-coating process.
To evaluate the intrinsic uniformity without the influence of such artifacts, we performed a statistical analysis of the resistance values in smaller square regions that do not contain any open or shorted junctions and are unaffected by these resist-thickness undulations, as shown in Fig.~\ref{figS:sdshape}(a).
These regions were selected to minimize the contribution from long-range spatial nonuniformities.
As shown in the histograms in Fig.~\ref{figS:sdshape}(b)--(d), the distributions are nearly symmetric, and the relative standard deviations are \SI{0.73}{\percent} for the \SI{4.5}{\milli\meter} square region with 961 junctions, \SI{0.63}{\percent} for the \SI{3.0}{\milli\meter} square region with 441 junctions, and \SI{0.50}{\percent} for the \SI{1.5}{\milli\meter} square region with 121 junctions.
Because variations in resist thickness exhibit long correlation lengths, the relative standard deviation systematically decreases as the region size is reduced.
This result indicates that, if a spatially uniform resist thickness can be achieved, the variation can potentially be suppressed to the level of \SI{0.50}{\percent}, as demonstrated in the \SI{1.5}{\milli\meter} square region.

In summary, the model-based analysis of the junction-geometry dependence revealed the origin of the critical-current variation in Dolan-bridge Josephson junctions.
Among the modeled sources of variation, the film-thickness-related contribution was identified as the dominant component in the present dataset.
Furthermore, the model suggests that a deposition angle of \SI{30}{\degree} for bilayer junctions minimizes the thickness-induced contribution, and this optimization was experimentally verified.
The remaining long-range spatial nonuniformity observed over the chip is likely associated with resist-thickness variations introduced during the spin-coating process, which can be further mitigated through improved resist-coating procedures.
This approach provides practical guidelines for scaling up superconducting quantum circuits based on Dolan-bridge \ce{Al/AlO_x/Al} junctions.
In addition, the model-based statistical framework developed in this work can be extended to quantify variations across different junction geometries and fabrication processes.

We also emphasize that the obtained level of uniformity is compatible with the requirements of large-scale integration of superconducting qubits, where variation in the critical current directly affects qubit frequency dispersion.
By applying the same optimization to wafer-scale fabrication, the uniformity can be further improved, potentially reducing the calibration overhead of multi-qubit systems.

\section*{Supplementary Material}
See the supplementary material for details of the derivation of Eq. (\ref{eqS:fNL}).

\section*{Author Declarations}
\subsection*{Conflict of interest}
The authors have no conflicts to disclose.

\section*{Data Availability}
The data that support the findings of this study are available from the corresponding author upon reasonable request.

\bibliographystyle{aipnum4-1}

\begin{thebibliography}{16}%
\makeatletter
\providecommand \@ifxundefined [1]{%
 \@ifx{#1\undefined}
}%
\providecommand \@ifnum [1]{%
 \ifnum #1\expandafter \@firstoftwo
 \else \expandafter \@secondoftwo
 \fi
}%
\providecommand \@ifx [1]{%
 \ifx #1\expandafter \@firstoftwo
 \else \expandafter \@secondoftwo
 \fi
}%
\providecommand \natexlab [1]{#1}%
\providecommand \enquote  [1]{``#1''}%
\providecommand \bibnamefont  [1]{#1}%
\providecommand \bibfnamefont [1]{#1}%
\providecommand \citenamefont [1]{#1}%
\providecommand \href@noop [0]{\@secondoftwo}%
\providecommand \href [0]{\begingroup \@sanitize@url \@href}%
\providecommand \@href[1]{\@@startlink{#1}\@@href}%
\providecommand \@@href[1]{\endgroup#1\@@endlink}%
\providecommand \@sanitize@url [0]{\catcode `\\12\catcode `\$12\catcode
  `\&12\catcode `\#12\catcode `\^12\catcode `\_12\catcode `\%12\relax}%
\providecommand \@@startlink[1]{}%
\providecommand \@@endlink[0]{}%
\providecommand \url  [0]{\begingroup\@sanitize@url \@url }%
\providecommand \@url [1]{\endgroup\@href {#1}{\urlprefix }}%
\providecommand \urlprefix  [0]{URL }%
\providecommand \Eprint [0]{\href }%
\providecommand \doibase [0]{http://dx.doi.org/}%
\providecommand \selectlanguage [0]{\@gobble}%
\providecommand \bibinfo  [0]{\@secondoftwo}%
\providecommand \bibfield  [0]{\@secondoftwo}%
\providecommand \translation [1]{[#1]}%
\providecommand \BibitemOpen [0]{}%
\providecommand \bibitemStop [0]{}%
\providecommand \bibitemNoStop [0]{.\EOS\space}%
\providecommand \EOS [0]{\spacefactor3000\relax}%
\providecommand \BibitemShut  [1]{\csname bibitem#1\endcsname}%
\let\auto@bib@innerbib\@empty
\bibitem [{\citenamefont {Dolan}(1977)}]{10.1063/1.89690}%
  \BibitemOpen
  \bibfield  {author} {\bibinfo {author} {\bibfnamefont {G.~J.}\ \bibnamefont
  {Dolan}},\ }\href {\doibase 10.1063/1.89690} {\bibfield  {journal} {\bibinfo
  {journal} {Applied Physics Letters}\ }\textbf {\bibinfo {volume} {31}},\
  \bibinfo {pages} {337} (\bibinfo {year} {1977})}
  \BibitemShut {NoStop}%
\bibitem [{\citenamefont {Potts}\ \emph {et~al.}(2001)\citenamefont {Potts},
  \citenamefont {Routley}, \citenamefont {Parker}, \citenamefont {Baumberg},\
  and\ \citenamefont {de~Groot}}]{Potts2001}%
  \BibitemOpen
  \bibfield  {author} {\bibinfo {author} {\bibfnamefont {A.}~\bibnamefont
  {Potts}}, \bibinfo {author} {\bibfnamefont {P.~R.}\ \bibnamefont {Routley}},
  \bibinfo {author} {\bibfnamefont {G.~J.}\ \bibnamefont {Parker}}, \bibinfo
  {author} {\bibfnamefont {J.~J.}\ \bibnamefont {Baumberg}}, \ and\ \bibinfo
  {author} {\bibfnamefont {P.~A.~J.}\ \bibnamefont {de~Groot}},\ }\href
  {\doibase 10.1023/A:1011279908265} {\bibfield  {journal} {\bibinfo  {journal}
  {Journal of Materials Science: Materials in Electronics}\ }\textbf {\bibinfo
  {volume} {12}},\ \bibinfo {pages} {289} (\bibinfo {year} {2001})}\BibitemShut
  {NoStop}%
\bibitem [{\citenamefont {Costache}\ \emph {et~al.}(2012)\citenamefont
  {Costache}, \citenamefont {Bridoux}, \citenamefont {Neumann},\ and\
  \citenamefont {Valenzuela}}]{10.1116/1.4722982}%
  \BibitemOpen
  \bibfield  {author} {\bibinfo {author} {\bibfnamefont {M.~V.}\ \bibnamefont
  {Costache}}, \bibinfo {author} {\bibfnamefont {G.}~\bibnamefont {Bridoux}},
  \bibinfo {author} {\bibfnamefont {I.}~\bibnamefont {Neumann}}, \ and\
  \bibinfo {author} {\bibfnamefont {S.~O.}\ \bibnamefont {Valenzuela}},\ }\href
  {\doibase 10.1116/1.4722982} {\bibfield  {journal} {\bibinfo  {journal}
  {Journal of Vacuum Science \& Technology B}\ }\textbf {\bibinfo {volume}
  {30}},\ \bibinfo {pages} {04E105} (\bibinfo {year} {2012})}
  \BibitemShut {NoStop}%
\bibitem [{\citenamefont {Nakamura}, \citenamefont {Pashkin},\ and\
  \citenamefont {Tsai}(1999)}]{Nakamura1999}%
  \BibitemOpen
  \bibfield  {author} {\bibinfo {author} {\bibfnamefont {Y.}~\bibnamefont
  {Nakamura}}, \bibinfo {author} {\bibfnamefont {Y.~A.}\ \bibnamefont
  {Pashkin}}, \ and\ \bibinfo {author} {\bibfnamefont {J.~S.}\ \bibnamefont
  {Tsai}},\ }\href {\doibase 10.1038/19718} {\bibfield  {journal} {\bibinfo
  {journal} {Nature}\ }\textbf {\bibinfo {volume} {398}},\ \bibinfo {pages}
  {786} (\bibinfo {year} {1999})}\BibitemShut {NoStop}%
\bibitem [{\citenamefont {Koch}\ \emph {et~al.}(2007)\citenamefont {Koch},
  \citenamefont {Yu}, \citenamefont {Gambetta}, \citenamefont {Houck},
  \citenamefont {Schuster}, \citenamefont {Majer}, \citenamefont {Blais},
  \citenamefont {Devoret}, \citenamefont {Girvin},\ and\ \citenamefont
  {Schoelkopf}}]{PhysRevA.76.042319}%
  \BibitemOpen
  \bibfield  {author} {\bibinfo {author} {\bibfnamefont {J.}~\bibnamefont
  {Koch}}, \bibinfo {author} {\bibfnamefont {T.~M.}\ \bibnamefont {Yu}},
  \bibinfo {author} {\bibfnamefont {J.}~\bibnamefont {Gambetta}}, \bibinfo
  {author} {\bibfnamefont {A.~A.}\ \bibnamefont {Houck}}, \bibinfo {author}
  {\bibfnamefont {D.~I.}\ \bibnamefont {Schuster}}, \bibinfo {author}
  {\bibfnamefont {J.}~\bibnamefont {Majer}}, \bibinfo {author} {\bibfnamefont
  {A.}~\bibnamefont {Blais}}, \bibinfo {author} {\bibfnamefont {M.~H.}\
  \bibnamefont {Devoret}}, \bibinfo {author} {\bibfnamefont {S.~M.}\
  \bibnamefont {Girvin}}, \ and\ \bibinfo {author} {\bibfnamefont {R.~J.}\
  \bibnamefont {Schoelkopf}},\ }\href {\doibase 10.1103/PhysRevA.76.042319}
  {\bibfield  {journal} {\bibinfo  {journal} {Phys. Rev. A}\ }\textbf {\bibinfo
  {volume} {76}},\ \bibinfo {pages} {042319} (\bibinfo {year}
  {2007})}\BibitemShut {NoStop}%
\bibitem [{\citenamefont {Arute}\ \emph {et~al.}(2019)\citenamefont {Arute},
  \citenamefont {Arya}, \citenamefont {Babbush}, \citenamefont {Bacon},
  \citenamefont {Bardin}, \citenamefont {Barends}, \citenamefont {Biswas},
  \citenamefont {Boixo}, \citenamefont {Brandao}, \citenamefont {Buell},
  \citenamefont {Burkett}, \citenamefont {Chen}, \citenamefont {Chen},
  \citenamefont {Chiaro}, \citenamefont {Collins}, \citenamefont {Courtney},
  \citenamefont {Dunsworth}, \citenamefont {Farhi}, \citenamefont {Foxen},
  \citenamefont {Fowler}, \citenamefont {Gidney}, \citenamefont {Giustina},
  \citenamefont {Graff}, \citenamefont {Guerin}, \citenamefont {Habegger},
  \citenamefont {Harrigan}, \citenamefont {Hartmann}, \citenamefont {Ho},
  \citenamefont {Hoffmann}, \citenamefont {Huang}, \citenamefont {Humble},
  \citenamefont {Isakov}, \citenamefont {Jeffrey}, \citenamefont {Jiang},
  \citenamefont {Kafri}, \citenamefont {Kechedzhi}, \citenamefont {Kelly},
  \citenamefont {Klimov}, \citenamefont {Knysh}, \citenamefont {Korotkov},
  \citenamefont {Kostritsa}, \citenamefont {Landhuis}, \citenamefont
  {Lindmark}, \citenamefont {Lucero}, \citenamefont {Lyakh}, \citenamefont
  {Mandra}, \citenamefont {McClean}, \citenamefont {McEwen}, \citenamefont
  {Megrant}, \citenamefont {Mi}, \citenamefont {Michielsen}, \citenamefont
  {Mohseni}, \citenamefont {Mutus}, \citenamefont {Naaman}, \citenamefont
  {Neeley}, \citenamefont {Neill}, \citenamefont {Niu}, \citenamefont {Ostby},
  \citenamefont {Petukhov}, \citenamefont {Platt}, \citenamefont {Quintana},
  \citenamefont {Rieffel}, \citenamefont {Roushan}, \citenamefont {Rubin},
  \citenamefont {Sank}, \citenamefont {Satzinger}, \citenamefont {Smelyanskiy},
  \citenamefont {Sung}, \citenamefont {Trevithick}, \citenamefont
  {Vainsencher}, \citenamefont {Villalonga}, \citenamefont {White},
  \citenamefont {Yao}, \citenamefont {Yeh}, \citenamefont {Zalcman},
  \citenamefont {Neven},\ and\ \citenamefont {Martinis}}]{Arute2019}%
  \BibitemOpen
  \bibfield  {author} {\bibinfo {author} {\bibfnamefont {F.}~\bibnamefont
  {Arute}}, \bibinfo {author} {\bibfnamefont {K.}~\bibnamefont {Arya}},
  \bibinfo {author} {\bibfnamefont {R.}~\bibnamefont {Babbush}}, \bibinfo
  {author} {\bibfnamefont {D.}~\bibnamefont {Bacon}}, \bibinfo {author}
  {\bibfnamefont {J.~C.}\ \bibnamefont {Bardin}}, \bibinfo {author}
  {\bibfnamefont {R.}~\bibnamefont {Barends}}, \bibinfo {author} {\bibfnamefont
  {R.}~\bibnamefont {Biswas}}, \bibinfo {author} {\bibfnamefont
  {S.}~\bibnamefont {Boixo}}, \bibinfo {author} {\bibfnamefont {F.~G. S.~L.}\
  \bibnamefont {Brandao}}, \bibinfo {author} {\bibfnamefont {D.~A.}\
  \bibnamefont {Buell}}, \bibinfo {author} {\bibfnamefont {B.}~\bibnamefont
  {Burkett}}, \bibinfo {author} {\bibfnamefont {Y.}~\bibnamefont {Chen}},
  \bibinfo {author} {\bibfnamefont {Z.}~\bibnamefont {Chen}}, \bibinfo {author}
  {\bibfnamefont {B.}~\bibnamefont {Chiaro}}, \bibinfo {author} {\bibfnamefont
  {R.}~\bibnamefont {Collins}}, \bibinfo {author} {\bibfnamefont
  {W.}~\bibnamefont {Courtney}}, \bibinfo {author} {\bibfnamefont
  {A.}~\bibnamefont {Dunsworth}}, \bibinfo {author} {\bibfnamefont
  {E.}~\bibnamefont {Farhi}}, \bibinfo {author} {\bibfnamefont
  {B.}~\bibnamefont {Foxen}}, \bibinfo {author} {\bibfnamefont
  {A.}~\bibnamefont {Fowler}}, \bibinfo {author} {\bibfnamefont
  {C.}~\bibnamefont {Gidney}}, \bibinfo {author} {\bibfnamefont
  {M.}~\bibnamefont {Giustina}}, \bibinfo {author} {\bibfnamefont
  {R.}~\bibnamefont {Graff}}, \bibinfo {author} {\bibfnamefont
  {K.}~\bibnamefont {Guerin}}, \bibinfo {author} {\bibfnamefont
  {S.}~\bibnamefont {Habegger}}, \bibinfo {author} {\bibfnamefont {M.~P.}\
  \bibnamefont {Harrigan}}, \bibinfo {author} {\bibfnamefont {M.~J.}\
  \bibnamefont {Hartmann}}, \bibinfo {author} {\bibfnamefont {A.}~\bibnamefont
  {Ho}}, \bibinfo {author} {\bibfnamefont {M.}~\bibnamefont {Hoffmann}},
  \bibinfo {author} {\bibfnamefont {T.}~\bibnamefont {Huang}}, \bibinfo
  {author} {\bibfnamefont {T.~S.}\ \bibnamefont {Humble}}, \bibinfo {author}
  {\bibfnamefont {S.~V.}\ \bibnamefont {Isakov}}, \bibinfo {author}
  {\bibfnamefont {E.}~\bibnamefont {Jeffrey}}, \bibinfo {author} {\bibfnamefont
  {Z.}~\bibnamefont {Jiang}}, \bibinfo {author} {\bibfnamefont
  {D.}~\bibnamefont {Kafri}}, \bibinfo {author} {\bibfnamefont
  {K.}~\bibnamefont {Kechedzhi}}, \bibinfo {author} {\bibfnamefont
  {J.}~\bibnamefont {Kelly}}, \bibinfo {author} {\bibfnamefont {P.~V.}\
  \bibnamefont {Klimov}}, \bibinfo {author} {\bibfnamefont {S.}~\bibnamefont
  {Knysh}}, \bibinfo {author} {\bibfnamefont {A.}~\bibnamefont {Korotkov}},
  \bibinfo {author} {\bibfnamefont {F.}~\bibnamefont {Kostritsa}}, \bibinfo
  {author} {\bibfnamefont {D.}~\bibnamefont {Landhuis}}, \bibinfo {author}
  {\bibfnamefont {M.}~\bibnamefont {Lindmark}}, \bibinfo {author}
  {\bibfnamefont {E.}~\bibnamefont {Lucero}}, \bibinfo {author} {\bibfnamefont
  {D.}~\bibnamefont {Lyakh}}, \bibinfo {author} {\bibfnamefont
  {S.}~\bibnamefont {Mandra}}, \bibinfo {author} {\bibfnamefont {J.~R.}\
  \bibnamefont {McClean}}, \bibinfo {author} {\bibfnamefont {M.}~\bibnamefont
  {McEwen}}, \bibinfo {author} {\bibfnamefont {A.}~\bibnamefont {Megrant}},
  \bibinfo {author} {\bibfnamefont {X.}~\bibnamefont {Mi}}, \bibinfo {author}
  {\bibfnamefont {K.}~\bibnamefont {Michielsen}}, \bibinfo {author}
  {\bibfnamefont {M.}~\bibnamefont {Mohseni}}, \bibinfo {author} {\bibfnamefont
  {J.}~\bibnamefont {Mutus}}, \bibinfo {author} {\bibfnamefont
  {O.}~\bibnamefont {Naaman}}, \bibinfo {author} {\bibfnamefont
  {M.}~\bibnamefont {Neeley}}, \bibinfo {author} {\bibfnamefont
  {C.}~\bibnamefont {Neill}}, \bibinfo {author} {\bibfnamefont {M.~Y.}\
  \bibnamefont {Niu}}, \bibinfo {author} {\bibfnamefont {E.}~\bibnamefont
  {Ostby}}, \bibinfo {author} {\bibfnamefont {A.}~\bibnamefont {Petukhov}},
  \bibinfo {author} {\bibfnamefont {J.~C.}\ \bibnamefont {Platt}}, \bibinfo
  {author} {\bibfnamefont {C.}~\bibnamefont {Quintana}}, \bibinfo {author}
  {\bibfnamefont {E.~G.}\ \bibnamefont {Rieffel}}, \bibinfo {author}
  {\bibfnamefont {P.}~\bibnamefont {Roushan}}, \bibinfo {author} {\bibfnamefont
  {N.~C.}\ \bibnamefont {Rubin}}, \bibinfo {author} {\bibfnamefont
  {D.}~\bibnamefont {Sank}}, \bibinfo {author} {\bibfnamefont {K.~J.}\
  \bibnamefont {Satzinger}}, \bibinfo {author} {\bibfnamefont {V.}~\bibnamefont
  {Smelyanskiy}}, \bibinfo {author} {\bibfnamefont {K.~J.}\ \bibnamefont
  {Sung}}, \bibinfo {author} {\bibfnamefont {M.~D.}\ \bibnamefont
  {Trevithick}}, \bibinfo {author} {\bibfnamefont {A.}~\bibnamefont
  {Vainsencher}}, \bibinfo {author} {\bibfnamefont {B.}~\bibnamefont
  {Villalonga}}, \bibinfo {author} {\bibfnamefont {T.}~\bibnamefont {White}},
  \bibinfo {author} {\bibfnamefont {Z.~J.}\ \bibnamefont {Yao}}, \bibinfo
  {author} {\bibfnamefont {P.}~\bibnamefont {Yeh}}, \bibinfo {author}
  {\bibfnamefont {A.}~\bibnamefont {Zalcman}}, \bibinfo {author} {\bibfnamefont
  {H.}~\bibnamefont {Neven}}, \ and\ \bibinfo {author} {\bibfnamefont {J.~M.}\
  \bibnamefont {Martinis}},\ }\href {\doibase 10.1038/s41586-019-1666-5}
  {\bibfield  {journal} {\bibinfo  {journal} {Nature}\ }\textbf {\bibinfo
  {volume} {574}},\ \bibinfo {pages} {505} (\bibinfo {year}
  {2019})}\BibitemShut {NoStop}%
\bibitem [{\citenamefont {Brink}\ \emph {et~al.}(2018)\citenamefont {Brink},
  \citenamefont {Chow}, \citenamefont {Hertzberg}, \citenamefont {Magesan},\
  and\ \citenamefont {Rosenblatt}}]{8614500}%
  \BibitemOpen
  \bibfield  {author} {\bibinfo {author} {\bibfnamefont {M.}~\bibnamefont
  {Brink}}, \bibinfo {author} {\bibfnamefont {J.~M.}\ \bibnamefont {Chow}},
  \bibinfo {author} {\bibfnamefont {J.}~\bibnamefont {Hertzberg}}, \bibinfo
  {author} {\bibfnamefont {E.}~\bibnamefont {Magesan}}, \ and\ \bibinfo
  {author} {\bibfnamefont {S.}~\bibnamefont {Rosenblatt}},\ }in\ \href
  {\doibase 10.1109/IEDM.2018.8614500} {\emph {\bibinfo {booktitle} {2018 IEEE
  International Electron Devices Meeting (IEDM)}}}\ (\bibinfo {year} {2018})\
  pp.\ \bibinfo {pages} {6.1.1--6.1.3}\BibitemShut {NoStop}%
\bibitem [{\citenamefont {White}\ \emph {et~al.}(2015)\citenamefont {White},
  \citenamefont {Mutus}, \citenamefont {Hoi}, \citenamefont {Barends},
  \citenamefont {Campbell}, \citenamefont {Chen}, \citenamefont {Chen},
  \citenamefont {Chiaro}, \citenamefont {Dunsworth}, \citenamefont {Jeffrey},
  \citenamefont {Kelly}, \citenamefont {Megrant}, \citenamefont {Neill},
  \citenamefont {O'Malley}, \citenamefont {Roushan}, \citenamefont {Sank},
  \citenamefont {Vainsencher}, \citenamefont {Wenner}, \citenamefont
  {Chaudhuri}, \citenamefont {Gao},\ and\ \citenamefont
  {Martinis}}]{10.1063/1.4922348}%
  \BibitemOpen
  \bibfield  {author} {\bibinfo {author} {\bibfnamefont {T.~C.}\ \bibnamefont
  {White}}, \bibinfo {author} {\bibfnamefont {J.~Y.}\ \bibnamefont {Mutus}},
  \bibinfo {author} {\bibfnamefont {I.-C.}\ \bibnamefont {Hoi}}, \bibinfo
  {author} {\bibfnamefont {R.}~\bibnamefont {Barends}}, \bibinfo {author}
  {\bibfnamefont {B.}~\bibnamefont {Campbell}}, \bibinfo {author}
  {\bibfnamefont {Y.}~\bibnamefont {Chen}}, \bibinfo {author} {\bibfnamefont
  {Z.}~\bibnamefont {Chen}}, \bibinfo {author} {\bibfnamefont {B.}~\bibnamefont
  {Chiaro}}, \bibinfo {author} {\bibfnamefont {A.}~\bibnamefont {Dunsworth}},
  \bibinfo {author} {\bibfnamefont {E.}~\bibnamefont {Jeffrey}}, \bibinfo
  {author} {\bibfnamefont {J.}~\bibnamefont {Kelly}}, \bibinfo {author}
  {\bibfnamefont {A.}~\bibnamefont {Megrant}}, \bibinfo {author} {\bibfnamefont
  {C.}~\bibnamefont {Neill}}, \bibinfo {author} {\bibfnamefont {P.~J.~J.}\
  \bibnamefont {O'Malley}}, \bibinfo {author} {\bibfnamefont {P.}~\bibnamefont
  {Roushan}}, \bibinfo {author} {\bibfnamefont {D.}~\bibnamefont {Sank}},
  \bibinfo {author} {\bibfnamefont {A.}~\bibnamefont {Vainsencher}}, \bibinfo
  {author} {\bibfnamefont {J.}~\bibnamefont {Wenner}}, \bibinfo {author}
  {\bibfnamefont {S.}~\bibnamefont {Chaudhuri}}, \bibinfo {author}
  {\bibfnamefont {J.}~\bibnamefont {Gao}}, \ and\ \bibinfo {author}
  {\bibfnamefont {J.~M.}\ \bibnamefont {Martinis}},\ }\href {\doibase
  10.1063/1.4922348} {\bibfield  {journal} {\bibinfo  {journal} {Applied
  Physics Letters}\ }\textbf {\bibinfo {volume} {106}},\ \bibinfo {pages}
  {242601} (\bibinfo {year} {2015})}
  \BibitemShut {NoStop}%
\bibitem [{\citenamefont {Macklin}\ \emph {et~al.}(2015)\citenamefont
  {Macklin}, \citenamefont {O’Brien}, \citenamefont {Hover}, \citenamefont
  {Schwartz}, \citenamefont {Bolkhovsky}, \citenamefont {Zhang}, \citenamefont
  {Oliver},\ and\ \citenamefont {Siddiqi}}]{doi:10.1126/science.aaa8525}%
  \BibitemOpen
  \bibfield  {author} {\bibinfo {author} {\bibfnamefont {C.}~\bibnamefont
  {Macklin}}, \bibinfo {author} {\bibfnamefont {K.}~\bibnamefont {O’Brien}},
  \bibinfo {author} {\bibfnamefont {D.}~\bibnamefont {Hover}}, \bibinfo
  {author} {\bibfnamefont {M.~E.}\ \bibnamefont {Schwartz}}, \bibinfo {author}
  {\bibfnamefont {V.}~\bibnamefont {Bolkhovsky}}, \bibinfo {author}
  {\bibfnamefont {X.}~\bibnamefont {Zhang}}, \bibinfo {author} {\bibfnamefont
  {W.~D.}\ \bibnamefont {Oliver}}, \ and\ \bibinfo {author} {\bibfnamefont
  {I.}~\bibnamefont {Siddiqi}},\ }\href {\doibase 10.1126/science.aaa8525}
  {\bibfield  {journal} {\bibinfo  {journal} {Science}\ }\textbf {\bibinfo
  {volume} {350}},\ \bibinfo {pages} {307} (\bibinfo {year} {2015})}
  \BibitemShut
  {NoStop}%
\bibitem [{\citenamefont {\'O~Peat\'ain}\ \emph {et~al.}(2023)\citenamefont
  {\'O~Peat\'ain}, \citenamefont {Dixon}, \citenamefont {Meeson}, \citenamefont
  {Williams}, \citenamefont {Kafanov},\ and\ \citenamefont
  {Pashkin}}]{Peatain_2023}%
  \BibitemOpen
  \bibfield  {author} {\bibinfo {author} {\bibfnamefont {S.}~\bibnamefont
  {\'O~Peat\'ain}}, \bibinfo {author} {\bibfnamefont {T.}~\bibnamefont
  {Dixon}}, \bibinfo {author} {\bibfnamefont {P.~J.}\ \bibnamefont {Meeson}},
  \bibinfo {author} {\bibfnamefont {J.~M.}\ \bibnamefont {Williams}}, \bibinfo
  {author} {\bibfnamefont {S.}~\bibnamefont {Kafanov}}, \ and\ \bibinfo
  {author} {\bibfnamefont {Y.~A.}\ \bibnamefont {Pashkin}},\ }\href {\doibase
  10.1088/1361-6668/acba4e} {\bibfield  {journal} {\bibinfo  {journal}
  {Superconductor Science and Technology}\ }\textbf {\bibinfo {volume} {36}},\
  \bibinfo {pages} {045017} (\bibinfo {year} {2023})}\BibitemShut {NoStop}%
\bibitem [{\citenamefont {Kreikebaum}\ \emph {et~al.}(2020)\citenamefont
  {Kreikebaum}, \citenamefont {O’Brien}, \citenamefont {Morvan},\ and\
  \citenamefont {Siddiqi}}]{Kreikebaum_2020}%
  \BibitemOpen
  \bibfield  {author} {\bibinfo {author} {\bibfnamefont {J.~M.}\ \bibnamefont
  {Kreikebaum}}, \bibinfo {author} {\bibfnamefont {K.~P.}\ \bibnamefont
  {O’Brien}}, \bibinfo {author} {\bibfnamefont {A.}~\bibnamefont {Morvan}}, \
  and\ \bibinfo {author} {\bibfnamefont {I.}~\bibnamefont {Siddiqi}},\ }\href
  {\doibase 10.1088/1361-6668/ab8617} {\bibfield  {journal} {\bibinfo
  {journal} {Superconductor Science and Technology}\ }\textbf {\bibinfo
  {volume} {33}},\ \bibinfo {pages} {06LT02} (\bibinfo {year}
  {2020})}\BibitemShut {NoStop}%
\bibitem [{\citenamefont {Takahashi}\ \emph {et~al.}(2022)\citenamefont
  {Takahashi}, \citenamefont {Kouma}, \citenamefont {Doi}, \citenamefont
  {Sato}, \citenamefont {Tamate},\ and\ \citenamefont
  {Nakamura}}]{Takahashi_2023}%
  \BibitemOpen
  \bibfield  {author} {\bibinfo {author} {\bibfnamefont {T.}~\bibnamefont
  {Takahashi}}, \bibinfo {author} {\bibfnamefont {N.}~\bibnamefont {Kouma}},
  \bibinfo {author} {\bibfnamefont {Y.}~\bibnamefont {Doi}}, \bibinfo {author}
  {\bibfnamefont {S.}~\bibnamefont {Sato}}, \bibinfo {author} {\bibfnamefont
  {S.}~\bibnamefont {Tamate}}, \ and\ \bibinfo {author} {\bibfnamefont
  {Y.}~\bibnamefont {Nakamura}},\ }\href {\doibase 10.35848/1347-4065/aca256}
  {\bibfield  {journal} {\bibinfo  {journal} {Japanese Journal of Applied
  Physics}\ }\textbf {\bibinfo {volume} {62}},\ \bibinfo {pages} {SC1002}
  (\bibinfo {year} {2022})}\BibitemShut {NoStop}%
\bibitem [{\citenamefont {Moskalev}\ \emph {et~al.}(2023)\citenamefont
  {Moskalev}, \citenamefont {Zikiy}, \citenamefont {Pishchimova}, \citenamefont
  {Ezenkova}, \citenamefont {Smirnov}, \citenamefont {Ivanov}, \citenamefont
  {Korshakov},\ and\ \citenamefont {Rodionov}}]{Moskalev2023}%
  \BibitemOpen
  \bibfield  {author} {\bibinfo {author} {\bibfnamefont {D.~O.}\ \bibnamefont
  {Moskalev}}, \bibinfo {author} {\bibfnamefont {E.~V.}\ \bibnamefont {Zikiy}},
  \bibinfo {author} {\bibfnamefont {A.~A.}\ \bibnamefont {Pishchimova}},
  \bibinfo {author} {\bibfnamefont {D.~A.}\ \bibnamefont {Ezenkova}}, \bibinfo
  {author} {\bibfnamefont {N.~S.}\ \bibnamefont {Smirnov}}, \bibinfo {author}
  {\bibfnamefont {A.~I.}\ \bibnamefont {Ivanov}}, \bibinfo {author}
  {\bibfnamefont {N.~D.}\ \bibnamefont {Korshakov}}, \ and\ \bibinfo {author}
  {\bibfnamefont {I.~A.}\ \bibnamefont {Rodionov}},\ }\href {\doibase
  10.1038/s41598-023-31003-1} {\bibfield  {journal} {\bibinfo  {journal}
  {Scientific Reports}\ }\textbf {\bibinfo {volume} {13}},\ \bibinfo {pages}
  {4174} (\bibinfo {year} {2023})}\BibitemShut {NoStop}%
\bibitem [{\citenamefont {Pishchimova}\ \emph {et~al.}(2023)\citenamefont
  {Pishchimova}, \citenamefont {Smirnov}, \citenamefont {Ezenkova},
  \citenamefont {Krivko}, \citenamefont {Zikiy}, \citenamefont {Moskalev},
  \citenamefont {Ivanov}, \citenamefont {Korshakov},\ and\ \citenamefont
  {Rodionov}}]{Pishchimova2023}%
  \BibitemOpen
  \bibfield  {author} {\bibinfo {author} {\bibfnamefont {A.~A.}\ \bibnamefont
  {Pishchimova}}, \bibinfo {author} {\bibfnamefont {N.~S.}\ \bibnamefont
  {Smirnov}}, \bibinfo {author} {\bibfnamefont {D.~A.}\ \bibnamefont
  {Ezenkova}}, \bibinfo {author} {\bibfnamefont {E.~A.}\ \bibnamefont
  {Krivko}}, \bibinfo {author} {\bibfnamefont {E.~V.}\ \bibnamefont {Zikiy}},
  \bibinfo {author} {\bibfnamefont {D.~O.}\ \bibnamefont {Moskalev}}, \bibinfo
  {author} {\bibfnamefont {A.~I.}\ \bibnamefont {Ivanov}}, \bibinfo {author}
  {\bibfnamefont {N.~D.}\ \bibnamefont {Korshakov}}, \ and\ \bibinfo {author}
  {\bibfnamefont {I.~A.}\ \bibnamefont {Rodionov}},\ }\href {\doibase
  10.1038/s41598-023-34051-9} {\bibfield  {journal} {\bibinfo  {journal}
  {Scientific Reports}\ }\textbf {\bibinfo {volume} {13}},\ \bibinfo {pages}
  {6772} (\bibinfo {year} {2023})}\BibitemShut {NoStop}%
\bibitem [{\citenamefont {Zheng}\ \emph {et~al.}(2023)\citenamefont {Zheng},
  \citenamefont {Li}, \citenamefont {Ding}, \citenamefont {Xiong},
  \citenamefont {Feng},\ and\ \citenamefont {Yang}}]{Zheng2023}%
  \BibitemOpen
  \bibfield  {author} {\bibinfo {author} {\bibfnamefont {Y.}~\bibnamefont
  {Zheng}}, \bibinfo {author} {\bibfnamefont {S.}~\bibnamefont {Li}}, \bibinfo
  {author} {\bibfnamefont {Z.}~\bibnamefont {Ding}}, \bibinfo {author}
  {\bibfnamefont {K.}~\bibnamefont {Xiong}}, \bibinfo {author} {\bibfnamefont
  {J.}~\bibnamefont {Feng}}, \ and\ \bibinfo {author} {\bibfnamefont
  {H.}~\bibnamefont {Yang}},\ }\href {\doibase 10.1038/s41598-023-39052-2}
  {\bibfield  {journal} {\bibinfo  {journal} {Scientific Reports}\ }\textbf
  {\bibinfo {volume} {13}},\ \bibinfo {pages} {11874} (\bibinfo {year}
  {2023})}\BibitemShut {NoStop}%
\bibitem [{\citenamefont {Ambegaokar}\ and\ \citenamefont
  {Baratoff}(1963)}]{PhysRevLett.10.486}%
  \BibitemOpen
  \bibfield  {author} {\bibinfo {author} {\bibfnamefont {V.}~\bibnamefont
  {Ambegaokar}}\ and\ \bibinfo {author} {\bibfnamefont {A.}~\bibnamefont
  {Baratoff}},\ }\href {\doibase 10.1103/PhysRevLett.10.486} {\bibfield
  {journal} {\bibinfo  {journal} {Phys. Rev. Lett.}\ }\textbf {\bibinfo
  {volume} {10}},\ \bibinfo {pages} {486} (\bibinfo {year} {1963})}\BibitemShut
  {NoStop}%
\end{thebibliography}

\end{document}


\title{
Supplementary Material\\
Geometric dependence of critical-current variation in \ce{Al/AlO_x/Al} Josephson junctions: a model-based analysis
}

\maketitle

\section{Derivation of the geometric factor $f^2(\theta,N_\mathrm{L})$ in Eq.~(13)}

This section derives the geometric factor $f^2(\theta,N_\mathrm{L})$ introduced in Eq.~(13) of the main text.
We consider a Dolan-bridge double-angle deposition with angles $\pm\theta$ (Fig.~1(a) in the main text).
Let $H$ be the vertical distance between the substrate plane and the intersection point of the two evaporation-beam center lines (a resist-geometry-dependent height parameter).
The thickness of the $n$-th deposited Al layer is denoted by $d_n$ ($n=1,\dots,N_\mathrm{L}$), and the junction width is $W$.

\subsection{Junction-interface area formed between layers $n$ and $n+1$}

In this geometry, the junction interface between the $n$-th and $(n+1)$-th layers consists of two contributions (see Fig.~1(a), right panel):
(i) the top-surface overlap and (ii) the sidewall-surface overlap.

After depositing the first $n$ layers, the effective vertical level relevant to the top overlap is shifted by $\sum_{k=1}^{n} d_k$.
The top-overlap length (projected in the lateral direction) is therefore
\begin{equation}
  \ell_n = 2\tan\theta \left( H - \sum_{k=1}^{n} d_k \right).
\end{equation}
The sidewall-overlap length associated with the $n$-th layer is
\begin{equation}
  s_n = \frac{d_n}{\cos\theta}.
\end{equation}
Thus, the junction-interface area created between layers $n$ and $n+1$ is
\begin{equation}
  S_{n,n+1} = W(\ell_n + s_n)
  = W\left[2\tan\theta \left( H - \sum_{k=1}^{n} d_k \right) + \frac{d_n}{\cos\theta}\right].
\end{equation}

For an $N_\mathrm{L}$-layer junction, the total junction-interface area is the sum over $n=1,\dots,N_\mathrm{L}-1$:
\begin{equation}
  S_{N_\mathrm{L}} = \sum_{n=1}^{N_\mathrm{L}-1} S_{n,n+1}.
\end{equation}

\subsection{Area fluctuation induced by thickness fluctuations}

We consider small thickness fluctuations $d_n \rightarrow d_n + \delta d_n$ and compute the induced area fluctuation
\begin{equation}
  \delta S_{N_\mathrm{L}} = \sum_{i=1}^{N_\mathrm{L}-1} \frac{\partial S_{N_\mathrm{L}}}{\partial d_i}\,\delta d_i.
\end{equation}
From the expression for $S_{n,n+1}$, the derivative with respect to $d_i$ receives
(i) a positive sidewall contribution from $s_i$ and (ii) negative top-overlap contributions because $d_i$ appears in $\ell_n$ for all $n\ge i$.
A straightforward differentiation yields
\begin{align}
  \frac{\partial S_{N_\mathrm{L}}}{\partial d_i}
  &= W\left[
      \frac{1}{\cos\theta}
      - \sum_{n=i}^{N_\mathrm{L}-1} 2\tan\theta
     \right] \nonumber\\
  &= W\left[
      \frac{1}{\cos\theta}
      - 2(N_\mathrm{L}-i)\tan\theta
     \right] \nonumber\\
  &= \frac{W}{\cos\theta}\left[
      1 - 2(N_\mathrm{L}-i)\sin\theta
     \right].
  \label{eq:dSd_di}
\end{align}

\subsection{Variance and closed-form expression}

Assuming that thickness fluctuations of different layers are uncorrelated and have the same variance,
\begin{equation}
  \langle \delta d_i\,\delta d_j\rangle = \delta_{ij}\,\langle(\delta d)^2\rangle,
\end{equation}
we obtain
\begin{align}
  \left\langle (\delta S_{N_\mathrm{L}})^2 \right\rangle
  &= \sum_{i=1}^{N_\mathrm{L}-1}
  \left(\frac{\partial S_{N_\mathrm{L}}}{\partial d_i}\right)^2
  \langle(\delta d)^2\rangle.
\end{align}
Using Eq.~\eqref{eq:dSd_di},
\begin{equation}
  \left(\frac{\partial S_{N_\mathrm{L}}}{\partial d_i}\right)^2
  = \frac{W^2}{\cos^2\theta}\left[1-2(N_\mathrm{L}-i)\sin\theta\right]^2.
\end{equation}

In the main text, the thickness-induced variance contribution is written as
\begin{equation}
  \mathrm{Var}(G_d) = c_d\, f^2(\theta,N_\mathrm{L})\,W^2,
\end{equation}
where $f^2(\theta,N_\mathrm{L})$ is defined (up to the proportionality constant absorbed into $c_d$) by
\begin{equation}
  f^2(\theta,N_\mathrm{L})
  = \frac{1}{W^2}\sum_{i=1}^{N_\mathrm{L}-1}
    \left(\frac{\partial S_{N_\mathrm{L}}}{\partial d_i}\right)^2
  = \frac{1}{\cos^2\theta}\sum_{i=1}^{N_\mathrm{L}-1}\left[1-2(N_\mathrm{L}-i)\sin\theta\right]^2.
\end{equation}

Let $m=N_\mathrm{L}-i$ so that $m=1,2,\dots,N_\mathrm{L}-1$. Then
\begin{equation}
  f^2(\theta,N_\mathrm{L})
  = \frac{1}{\cos^2\theta}\sum_{m=1}^{N_\mathrm{L}-1}\left(1-2m\sin\theta\right)^2
  = \frac{1}{\cos^2\theta}\sum_{m=1}^{N_\mathrm{L}-1}\left(1 - 4m\sin\theta + 4m^2\sin^2\theta\right).
\end{equation}
Using
\begin{equation}
  \sum_{m=1}^{N_\mathrm{L}-1} 1 = N_\mathrm{L}-1,\quad
  \sum_{m=1}^{N_\mathrm{L}-1} m = \frac{(N_\mathrm{L}-1)N_\mathrm{L}}{2},\quad
  \sum_{m=1}^{N_\mathrm{L}-1} m^2 = \frac{(N_\mathrm{L}-1)N_\mathrm{L}(2N_\mathrm{L}-1)}{6},
\end{equation}
we obtain the closed form
\begin{align}
  f^2(\theta,N_\mathrm{L})
  &= \frac{N_\mathrm{L}-1}{\cos^2\theta}
  \left[
    1 - 2N_\mathrm{L}\sin\theta + \frac{2N_\mathrm{L}(2N_\mathrm{L}-1)}{3}\sin^2\theta
  \right] \nonumber\\
  &= \frac{N_\mathrm{L}-1}{\cos^2\theta}
  \left[
    2N_\mathrm{L}\sin\theta\left(\frac{2N_\mathrm{L}-1}{3}\sin\theta - 1\right) + 1
  \right],
\end{align}
which reproduces Eq.~(13) in the main text.

\subsection{Special case: bilayer junction ($N_\mathrm{L}=2$)}

For $N_\mathrm{L}=2$, the above expression reduces to
\begin{equation}
  f^2(\theta,2)
  = \frac{1}{\cos^2\theta}\left(1-2\sin\theta\right)^2
  = \left(\frac{1}{\cos\theta} - 2\tan\theta\right)^2,
\end{equation}
which corresponds to Eq.~(14) in the main text.